\documentclass[namedreferences]{SolarPhysics}
\usepackage[optionalrh,solaenum,natbib]{spr-sola-addons} % For Solar Physics 
\usepackage{graphicx}                    % For eps figures, newer & more powerfull
\usepackage{color}                       % For color text: \color command
\usepackage{url}                         % For breaking URLs easily trough lines
\usepackage[pdfborder={0 0 0 },urlcolor=blue,breaklinks]{hyperref} 
\ifx \doiurl  \undefined \def \doiurl#1{\href{http://dx.doi.org/#1}{\url{#1}}}\fi
\ifx \adsurl  \undefined \def \adsurl#1{\href{http://adsabs.harvard.edu/abs/#1}{\url{#1}}}\fi

\begin{document}

\begin{article}

\begin{opening}

\title{Magnetic Field Strength and Inclination in the Penumbral Fine-Structure}

%%%%%%%%%%%%%%%%%%%%%%%%%%%%%%%%%%%%%%%%%%%%%%%%%%%
%% Authors Names
%
\author{E.~\surname{Wiehr}
              }

%%%%%%%%%%%%%%%%%%%%%%%%%%%%%%%%%%%%%%%%%%%%%%%%%%%
%% Runningheads
% 
\runningauthor{E. Wiehr}
\runningtitle{Magnetic Field Strength and Inclination in Penumbrae}

%%%%%%%%%%%%%%%%%%%%%%%%%%%%%%%%%%%%%%%%%%%%%%%%%%%
%% Affiliations 
%
  \institute{Institut f\"ur Astrophysik, G\"ottingen, Germany,
               email: \href{mailto:ewiehr@astro.physik.uni-goettingen.de} 
                          {ewiehr@astro.physik.uni-goettingen.de}\\
             }

%%%%%%%%%%%%%%%%%%%%%%%%%%%%%%%%%%%%%%%%%%%%%%%%%%%
%%% Abstract 
\begin{abstract}

The uncertainty about a possible correlation between strength and inclination 
of the magnetic field and the continuum intensity of the sunspot penumbral
fine-structure has been removed by a detailed analysis of a spatially very
high resolved spectrum:  the darker, lengthy penumbral lanes host a 10\% 
stronger and up to $30^o$ flatter magnetic field as compared to the bright 
penumbral locations. This finding does not only result from the high spatial 
resolution achived but also from the choice of the Fe\,I\,6842.7 line, which 
obtains its essential contribution from those deep layers near the continuum 
intensity level where the penumbral structure is seen. The almost perfect 
correlation establishes that the penumbral structure is formed by two 
magnetic components, which mainly differ by their field inclination and less
by their field strength. The discrepancy with results from other Zeeman lines, 
as e.g. Fe\,I\,6302.5, indicates a different field structure above the white 
light penumbral layers.

\end{abstract}

%%%%%%%%%%%%%%%%%%%%%%%%%%%%%%%%%%%%%%%%%%%%%%%%%%%
%% Keywords
%

\end{opening}
%-------------------------------------------------

%%%%%%%%%%%%%%%%%%%%%%%%%%%%%%%%%%%%%%%%%%%%%%%%%%%
%% Sections
%
\section{Introduction}%\label{s:?} 

Sunspot penumbrae are highly structured with bright and dark 'fibrils' 
extending quasi-radially from the umbra to the surrounding photosphere. 
Their widths are too small to be spatially resolved in Earth-bound 
observations which are degraded by turbulence in the Earth's atmosphere 
(seeing). Although modern image processing can eliminate almost entirely 
this influence, the finally restored two-dimensional images with a spatial 
resolution at the limits of the solar telescope used, e.g. $\sim$200~km on 
the sun, contain only few information such as the temperature (using the 
Planck law; S\"utterlin \& Wiehr 1998). This temperature structure is commonly 
assumed to be caused by the magnetic field, which has so far not been measured 
at a spatial resolution near that of processed white light images, since a 
spectrograph and polarization optics are required. First attempts to remove 
influences of seeing from spectra (Keller \& Johannesson 1995, S\"utterlin 
\& Wiehr 2000) gave an improved spatial resolution for line intensity profiles, 
however, not for Zeeman polarization.

It is thus rather astonishing that it was possible to deduce from spectra with 
moderate spatial resolution `flatter and stronger magnetic fields in the 
dark structure' of sunspot penumbrae as claimed by Beckers \& Schr\"oter (1968).
Careful inspection of their photographic spectra, however, gives strong 
indication that they were referring to pronounced penumbral 'disturbances' 
rather than to smallest penumbral structures. It, indeed, was not until 1990 
that the spatial variation of the field {\it inclination} has been measured
independently by Lites, Scharmer, Skumanich (1990), Degenhardt \& Wiehr
(1991), and Schmidt et al. (1992).

However, the question of a possible relation with the field {\it strength} 
remained a subject of considerable controversy: Degenhardt \& Wiehr (1991), 
Lites et al. (1993), Stanchfield, Thomas, Lites (1997) found the steeper fields 
to be stronger, contradicting the 1968 finding. Even the question whether spatial 
variations of the penumbral field strength do exist at all, is still open: the 
observations span results between 400\,Gs (Lites, Scharmer, Skumanich 1990),
200\,Gs (Solanki, R\"uedi, Livingston 1992), and no spatial field variations
(Schmidt et al. 1992).

A systematic relation with the continuum brightness has hitherto not been found.
Although Schmidt et al. (1992) and Hofmann et al. (1994) report more horizontal 
fields in the dark structure (in agreement with Beckers \& Schr\"oter 1968), 
recently Bernasconi et al (1998) find weaker and more horizontal fields in the 
brighter structure. Lites, Scharmer, Skumanich (1990) say that `variations of 
the field strength are much less dramatic than those of the continuum intensity'. 

\section{Observations}
 
As long as (speckle) reconstructed data of the penumbral Zeeman polarization 
is not available, only frame-selected slit spectra may help to solve the 
persistent controversy. In addition, a spectral line is required which
probes those deep layers where the penumbral continuum structure does exist.
This is most reasonably fulfilled by Fe\,I\,6842.7\,\AA{} which has 
the same Land\'e factor (g=2.5) as the often used Fe\,I\,6302.5 line, but its 
larger wavelength yields a 1.09 times higher ratio of Zeeman splitting and 
line width $\Delta \lambda_Z/\Delta \lambda_D$. It therefore gives for a field 
strength of only 2000\,Gs already a separation of the $\sigma$-components of 
$2\Delta\lambda_Z=220$\,m\AA{} (Degenhardt \& Wiehr 1991) and is thus totally 
split in penumbrae (Fig.\,1; see also Wiehr 1999). This allows to directly
determine the field strength; and the amount of circular polarization 
(Stokes-V maximum) is then purely determined by the magnetic field inclination 
cos$\gamma$ (with respect to the line-of-sight).

Stokes spectra of highest spatial resolution were obtained on 20 August, 
1999, with the evacuated 45\,cm Gregory-Coud\'e solar telescope on Tenerife 
(Schr\"oter, Soltau, Wiehr 1985) observing a sunspot at $\vartheta \approx15^o$. 
The spectrograph slit of correspondingly 0.75\,arcsec width (i.e. $\approx500$\,km 
on the solar surface) intersects the penumbral fine-structure almost 
perpendicularly. Behind the slit, a quarter-wave plate followed by a calcite 
block is used to analyze the circularly polarized light from the longitudinal 
Zeeman effect. The 0.15\,sec integration of the CCD-camera gives penumbral 
continuum intensities of 4000\,counts. 

\eject

During a period of exceptionally good seeing, one spectrum among a series of 
bursts achieves nearly the spatial resolution limit of telescope plus spectrograph. 
It is processed in the usual manner, i.e. an elimination of the dark and the flat 
matrices, and the subtraction of the simultaneously taken left and right handed 
polarized sub-spectra. The resulting Stokes-V is shown in figure\,1 where the 
smallest spatial structure has a half-width of 0.7\,arcsec (see also Fig.\,4).

%% Figure 
\begin{figure}[ht] 
\centerline{\includegraphics[width=\textwidth,clip=]{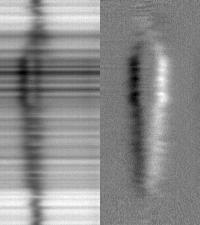}}
\caption{Blue $\sigma$-component of the Fe~6842.7 line ({\it left panel}) and 
circular Zeeman polarization Stokes-V ({\it right panel}) through a sunspot 
penumbra at $\vartheta \approx15^o$. The total spatial extension (slit length) 
amounts to 70 arcsec; the wavelength range covers 0.85 \AA{}.}
%\label{fig:1}
\end{figure}

%% Figure 
\begin{figure} 
\centerline{\includegraphics[width=\textwidth,clip=]{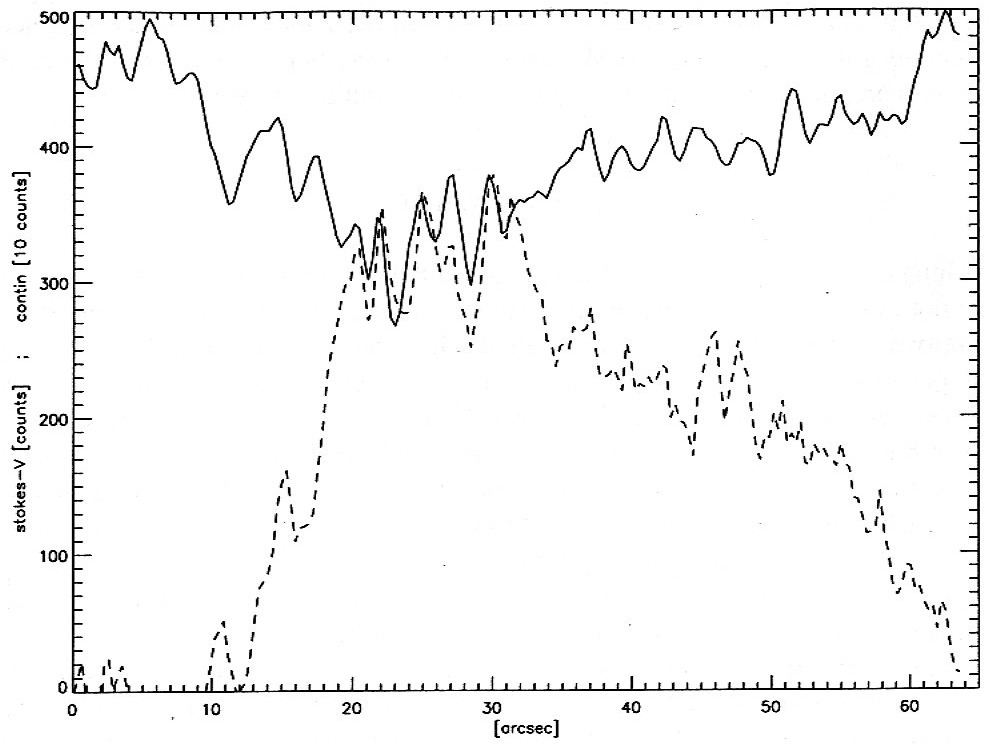}}
\caption{Continuum intensity ({\it solid line}) and total circular Zeeman 
polarization of Fe\,I\,6842.7 ({\it dashed line}) along a spatial cut 
(defined by the spectrograph slit) through a sunspot penumbra: dark 
penumbral continuum structures (i.e. local minima in the upper curve) 
are locations of smaller circular polarization, thus weaker and more 
horizontal magnetic field.}
%\label{fig:2}
\end{figure}

\section{\bf Results and their discussion} 

In order to obtain the spatial variation of intensity and total polarization with low 
noise, the continuum intensity is averaged over 0.5\AA{} (between two absorption 
lines), and the absolute amount of circular polarization is integrated over 0.5\AA{}, 
covering both Zeeman components. The resulting scans in figure\,2 show {\it larger 
circular Zeeman polarization in the brighter penumbral structures}.

Figure\,3 gives examples of Stokes-V profiles from three spatial locations at 21.2'', 
22.0'', and 23.0'' in figure\,2, representing one bright continuum streak and its two 
dark neighbors. The small differences of the Zeeman splitting indicate a {\it small 
spatial variation of the magnetic field strength}. The Stokes-V extrema indicate a
{\it stronger variation of the field inclination}. The scans show that the rms noise 
is of the order of $0.03\cdot I_{cont}$.

For a detailed investigation, the Zeeman components are fitted with Gaussian 
profiles at locations of complete splitting (i.e. 18''--33'' in figure\,2). Here, 
the wavelength separation of the polarization extrema, $2\Delta\lambda_Z$, is 
linearly related with the field strength, $B$; the Stokes-V amplitude then being 
a measure for the field inclination, cos$\gamma$. Figure\,3 gives the spatial 
variation of the total Stokes-V amplitude, $0.5\,(V^+-V^-)$, together with the 
spatial variation of $2\Delta\lambda_Z$ and the continuum intensity. It clearly 
shows that {\it less inclined and stronger fields are located in the dark 
penumbral lanes}. The Stokes-V variation can not result from the intensity 
variations alone since these are considerably smaller.

At locations in the outer penumbra, where the field strength is not high enough 
to yield a total line splitting, (i.e. outside the range 18''--33'' in figure\,2), 
the Stokes-V amplitude is proportional to $B\cdot cos\gamma$ thus not allowing
a separation of $B$ and $\gamma$. We may, however, reasonably extend the results 
found for locations with total splitting (figure\,4) also for the remaining spatial
locations, i.e., higher field strength in dark structures. In that case, the 
positive correlation of Stokes-V with the continuum intensity through the {\it whole} 
penumbra (cf., figure\,2) indicates that the {\it spatial variation of the inclination 
generally exceeds that of the field strength}. If, instead, the strength dominated, 
one would observe larger Stokes-V in dark rather than in bright continuum streaks, 
i.e. an anti-relation between Stokes-V and continuum, -- which is not observed.

%% Figure 
\begin{figure}
\centerline{\includegraphics[width=\textwidth,clip=]{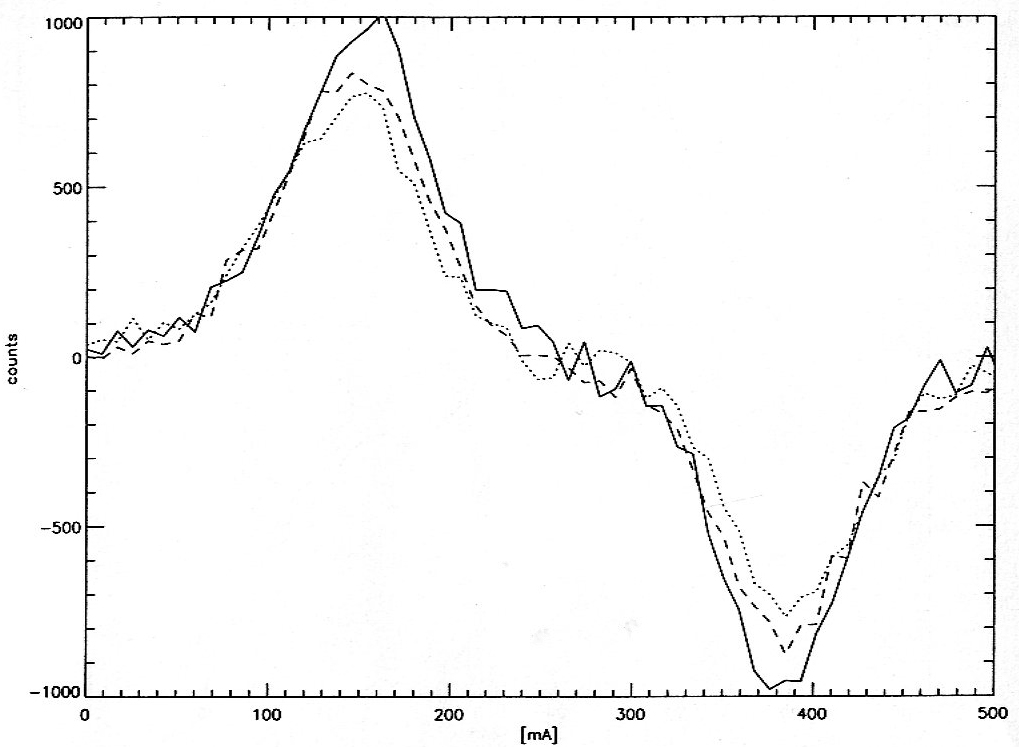}}
\caption{Stokes-V profiles of Fe\,I\,6842.7 in a bright continuum streak ({\it solid 
line}) located at  22.0'' in Figs.\,2 and 4 together those from its two dark neighbors 
at 21.2'' and 23.0'' ({\it dashed and dotted lines}). The bright continuum structure 
shows a slightly smaller Zeeman splitting but larger polarization maxima.}
%\label{fig:3}
\end{figure}

Particularly this latter finding demonstrates that the penumbral structure is mostly 
produced by a spatial variation of the {\it inclination} rather than of the strength 
of the magnetic field, which varies only by about 10\%. This picture of a sunspot 
penumbra nicely fits model calculations by Schlichenmaier, Jahn, Schmidt (1998)
proposing rising flux tubes which produce the locally steeper fields in the brighter 
penumbral structures. However, this model predicts that the latter are invisible 
since optically thick; the present results then suggest to reconsider that 
model. 

%% Figure 
%
\begin{figure}
\centerline{\includegraphics[width=1.05\textwidth,clip=]{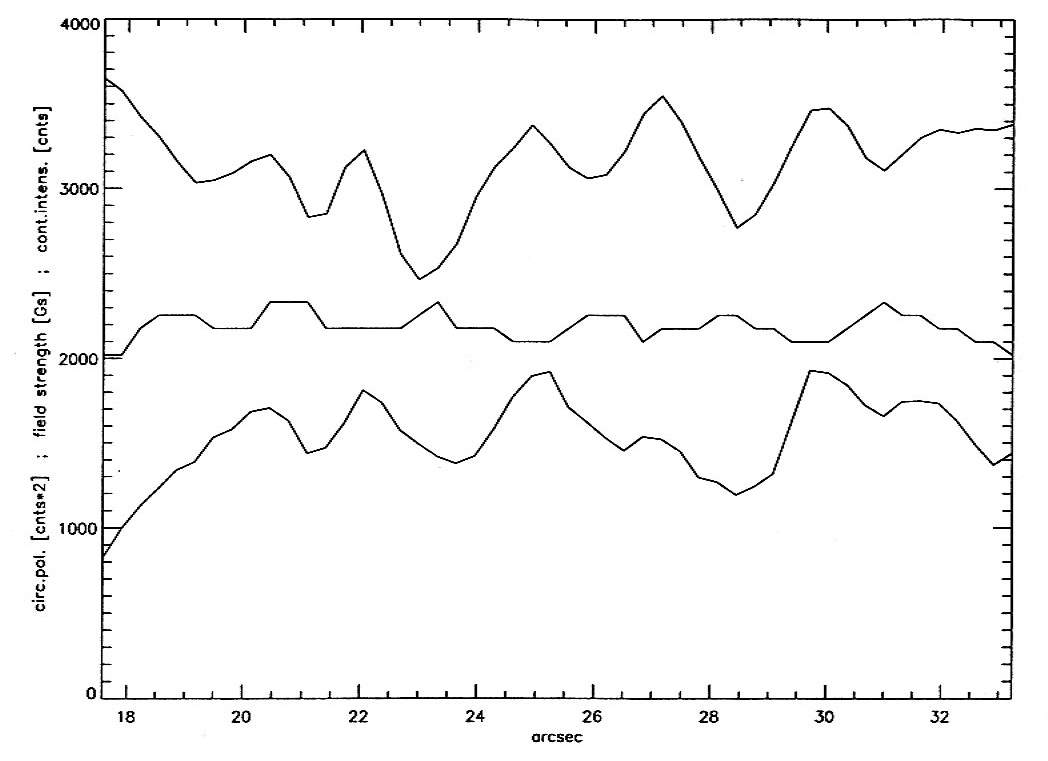}}
\caption{Close-up of figure\,2 at the locations 18''--33'' where the Fe\,I\,6842.7 
line is totally split thus yielding the magnetic field strength ({\it middle curve}) 
which is compared with the continuum intensity ({\it upper curve}) and Stokes-V 
({\it lower curve}). Dark penumbral lanes (minima in the upper curve) contain 
stronger and flatter magnetic fields.}
%\label{fig:4}
\end{figure}

\section{\bf Conclusion}

An important condition for the above findings is the total Zeeman splitting
of Fe\,6842.7\AA{} in penumbrae which allows the separation of strength and 
inclination of the magnetic field. This is not the case for the widely used 
Fe\,6302.5\AA{} (and Fe\,6173), which are not completely split by a penumbral 
magnetic field (Wiehr 1999). For such an incomplete splitting, a separation 
of field strength and inclination would require model calculations, which are 
potentially ambiguous since the detailed conditions in the various bright and 
dark penumbral structures are unknown. 

However, also at locations of complete Zeeman splitting (cf., figure\,3), a 
quantitative determination of the field inclination from the Stokes-V signal 
requires a knowledge of the model atmosphere in the respective bright or 
dark penumbral structure. Application of a mean penumbral model (Kjeldseth-Moe 
\& Maltby 1969) yields fluctuations $\Delta cos\gamma\le30^o$. But such a 
use of a mean atmosphere ignores that the line depth of Fe\,6842.7\AA{}, 
and thus Stokes-V, increases in hotter structures; -- this in addition to 
its intensity dependence. An interpretation of penumbral Stokes signals with 
a mean model can thus even hide essential information about the magnetic 
field (Wiehr 1974; section-6) and may well affect apparent 'variations' 
of the field inclination.

Even the application of separate models for bright and dark lanes 
(Kjeldseth-Moe \& Maltby 1974) will not help, since these are based on `typical' 
intensities of bright and dark penumbral structures with I$^b$=0.95 and 
I$^d$=0.61 (from Muller 1973), which have been contested by Grossmann-Doerth 
and Schmidt (1981) who find a single peaked histogram for penumbral intensities 
rather than a double peaked one. The latter reflects the fact that the penumbral 
structure is not defined in absolute brightness but only relative to its 
neighborhood, since dark structures do exist, which are even brighter than 
faintest bright structures in a penumbra (Wiehr et al. 1984). The different 
atmospheric conditions in individual penumbral structures include the 
possibility that observed variations of Stokes-V may be produced by their 
different atmospheres rather than by the actual magnetic field inclination.

Apart from this intrinsic complication, the present finding that dark penumbral 
structures host stronger and flatter fields essentially results from the fact 
that {\it the Fe\,6843 line used probes deep layers being close to those where 
the penumbral continuum structure occurs}. The absence of such a correlation 
in former data may essentially be due to the use of absorption lines, which obtain 
considerable contribution from layers above the penumbral white light structure.
The widely used Fe\,6303 line has an almost 2.5 times larger equivalent width, 
$W_{\lambda}$, and is not located on the linear part of the curve-of-growth but 
already saturates. It originates almost 200\,km above Fe\,6843 and hardly 
probes those deep layers, which host magnetic field and Evershed flow. This is 
also indicated in recent investigations by Schlichenmaier \& Schmidt (2000). 

The largely differing results deduced from lines, which obtain considerable 
contribution at larger heights, indicate that those higher layers contain 
different fields (and flows). In particular, the significantly smaller mean 
field strength obtained from lines probing higher layers (Wiehr 1999) may hardly
be explained by a simple field gradient. I like to suggest that only those
(inner) bright penumbral structures, which move toward the umbra, are the rising 
flux-tubes in the theoretical picture by Schlichenmaier, Jahn, Schmidt (1998).
Other (more outer) bright structures, which do not show a similarly strong 
radial motion in white light images, might represent the foot-points of a
(fainter) `background field' reaching high layers and affecting the
`super-penumbra' visible in chromospheric lines.
            
%%%%%%%%%%%%%%%%%%%%%%%%%%%%%%%%%%%%%%%%%%%%%%%%%%%%%%%%%%%%%%%%%%%%%%%%%%%
%% Acknowledgments
%
\begin{acks} 
The Gregory-Coud\'e Telescope on Tenerife is operated by the University 
of G\"ottingen at the Spanish 'Observatorio del Teide' of the Instituto 
de Astrofisica de Canarias.
\end{acks}
%%% %%%%%%%%%%%%%%%%%%%%%%%%%%%%%%%%%%%%%%%%%%%%%%%%%%%%%%%%%%%

\eject

%% Bibliography
%
% Using BibTeX
%
\bibliographystyle{spr-mp-sola}
% %\bibliographystyle{spr-mp-sola-cnd} %% Alternative style: no title, no concluding page
%\bibliography{helium}  
%
%\end{article} 
%\end{document}

% Without BibTeX 

\end{article} 
\end{document}